\documentstyle[11pt,dunk2001_asp,twoside,epsf]{article}
\def\lesssim{\mathrel{\hbox{\rlap{\hbox{\lower4pt\hbox{$\sim$}}}\hbox{$<$}}}}
\def\gtrsim{\mathrel{\hbox{\rlap{\hbox{\lower4pt\hbox{$\sim$}}}\hbox{$>$}}}}
\def\edcomment#1{\iffalse\marginpar{\raggedright\sl#1\/}\else\relax\fi}
\reversemarginpar
\begin{document}
\title{SAURON: An Innovative Look at Early-Type Galaxies}
\author{M.\ Bureau, M.\ Cappellari, Y.\ Copin, E.K.\ Verolme, P.T.\ de Zeeuw}
\affil{Sterrewacht Leiden, Niels Bohrweg~2, 2333~CA Leiden, Netherlands}
\author{R.\ Bacon, E.\ Emsellem}
\affil{CRAL, 9~Avenue Charles-Andr\'{e}, 69230 Saint-Genis-Laval, France}
\author{R.L.\ Davies, H.\ Kuntschner, R.\ McDermid}
\affil{Physics Department, University of Durham, South Road,
       Durham DH1~3LE, United Kingdom}
\author{B.W.\ Miller}
\affil{Gemini Observatory, Casilla~603, La Serena, Chile}
\author{R.F.\ Peletier}
\affil{Department of Physics and Astronomy, University of Nottingham,
  University Park, Nottingham NG7~2RD, United Kingdom}

\begin{abstract}
  A summary of the {\tt SAURON} project and its current status is presented.
  {\tt SAURON} is a panoramic integral-field spectrograph designed to study
  the stellar kinematics, gaseous kinematics, and stellar populations of
  spheroids. Here, the sample of galaxies and its properties are described.
  The instrument is detailed and its capabilities illustrated through
  observational examples. These includes results on the structure of central
  stellar disks, the kinematics and ionization state of gaseous disks, and the
  stellar populations of galaxies with decoupled cores.
\end{abstract}

\section{Introduction}

The physical properties of early-type galaxies correlate with luminosity and
environment. The morphology-density relation shows that ellipticals and
lenticular galaxies are much more common in clusters than in regions of lower
local density (Dressler 1980). Giant ellipticals ($M_B\lesssim-20.5$) are red,
have a high metal content, often have boxy isophotes and shallow cusps, and
are supported by anisotropic velocity distributions, associated with triaxial
shapes (e.g.\ de Zeeuw \& Franx 1991; Faber et al.\ 1997).  Lower-luminosity
systems ($M_B\gtrsim-20.5$) are bluer, less metal-rich, have disky isophotes
and steep cusps, and are flattened by rotation, suggesting nearly oblate
shapes (Davies et al.\ 1983; Bender \& Nieto 1990). The mass of the central
black hole in spheroids also correlates with the central velocity dispersion
(e.g.\ Gebhardt et al.\ 2000; Ferrarese \& Merritt 2000).

It is unclear to what extent these properties and the correlations between
them were acquired at the epoch of galaxy formation or result from subsequent
dynamical evolution. Key questions to which {\tt SAURON} hopes to provide
answers include: What is the distribution of intrinsic shapes, tumbling
speeds, and internal orbital structure among early-type galaxies? How do these
depend on total luminosity and environment? What is the shape and extent of
dark halos? What is the dynamical importance of central black holes? What is
the distribution of metals, and what is the relation between the kinematics of
stars (and gas), the local metal enrichment, and the star formation history?

Progress towards answering these questions requires a systematic investigation
of the kinematics and line-strengths of a representative sample of early-type
galaxies. The intrinsic shape, internal orbital structure, and radial
dependence of the mass-to-light ratio are constrained by the stellar and gas
kinematics (e.g.\ van der Marel \& Franx 1993; Cretton, Rix, \& de Zeeuw 2000);
the age and metallicity of the stellar populations by the absorption
line-strengths (Gonzalez 1993; Davies, Sadler, \& Peletier 1993). The {\tt
  SAURON} project will provide all of these data, and more, for a large and
well-defined sample of objects.\looseness=-2

\section{The Instrument}

Long-slit spectroscopy along a few position angles is insufficient to map the
rich internal kinematics of early-type galaxies (e.g.\ Statler 1991, 1994). We
thus built {\tt SAURON} ({\tt S}pectral {\tt A}real {\tt U}nit for {\tt
  R}esearch on {\tt O}ptical {\tt N}ebulae), a panoramic integral-field
spectrograph optimized for studies of the large-scale kinematics and stellar
populations of spheroids (Bacon et al.\ 2001, hereafter Paper~I). {\tt SAURON}
uses a lenslet array and is based on the {\tt TIGER} concept (Bacon et al.\
1995). In its low-resolution (LR) mode, it has a $41\arcsec\times33\arcsec$
field-of-view sampled with $0\farcs94\times0\farcs94$ lenslets, 100\%
coverage, and high throughput. In high-resolution (HR) mode, the field-of-view
is $11\arcsec\times9\arcsec$ sampled at $0\farcs27\times0\farcs27$. {\tt
  SAURON} simultaneously provides 1577 spectra over the wavelength range
4810--5350~\AA, 146 of which are used for sky subtraction. Stellar kinematic
information is derived from the Mg{\it b} triplet and the Fe lines; the
[OIII], H$\beta$, and [NI] emission lines provide the morphology, kinematics,
and ionization state of the ionized gas.  The Mg{\it b}, H$\beta$, and Fe5270
absorption lines are sensitive to the age and metallicity of the stellar
populations. The main characteristics of {\tt SAURON} are listed in Table~1.
Paper~I provides a full description of its design, construction, and of the
extensive data reduction software we developed. A pipeline called {\tt
  PALANTIR} is also described.

\section{The Sample}

Observing any complete sample which spans a wide range of properties is costly
in telescope time, even with {\tt SAURON}. We therefore constructed a {\it
  representative} sample of nearby ellipticals, lenticulars, and early-type
bulges, as free of biases as possible, but ensuring the existence of
complementary data. We also target some objects with known decoupled
kinematics (e.g.\ Davies et al.\ 2001). We will combine the {\tt SAURON}
observations with high-spatial resolution spectroscopy of the nuclei, mainly
from CFHT/{\tt OASIS} and HST/{\tt STIS}, and interpret them through dynamical
and stellar population modeling.

\begin{table}
\caption{{\tt SAURON}'s Main Characteristics}
\smallskip
\begin{tabular}{lcc}
\tableline
\tableline
\noalign{\smallskip}
 Property \phantom{XXXXXXXXXXXXXXXXXX}& \multicolumn{2}{c}{Mode} \\
      &\phantom{XXX} LR\phantom{XXX} &\phantom{XXX} HR\phantom{XXX}\\
\noalign{\smallskip}
\tableline
\noalign{\smallskip}
 Projected size of lenslet & 0\farcs94 & 0\farcs27 \\
 Field-of-view & $41\arcsec\times33\arcsec$ & $11\arcsec\times9\arcsec$ \\
 Spectral resolution (FWHM) & 3.6~\AA & 2.8~\AA \\
 Wavelength coverage & \multicolumn{2}{c}{4810--5350~\AA} \\
 Number of object lenslets & \multicolumn{2}{c}{1431} \\
 Number of sky lenslets & \multicolumn{2}{c}{146} \\
 Grism  & \multicolumn{2}{c}{514~lines~mm$^{-1}$} \\
 Spectral sampling & 1.1~\AA~pix$^{-1}$ & 0.9~\AA~pix$^{-1}$ \\
 Instrumental dispersion ($\sigma$) & 90~km~s$^{-1}$ & 70~km~s$^{-1}$  \\
 Spectra separation/PSF ratio & 1.4 & 2.3 \\
 Important spectral features & \multicolumn{2}{c}{H$\beta$,
         [OIII], Mg$_b$, FeI, [NI]} \\
 Calibration lamps & \multicolumn{2}{c}{Ne, Ar} \\
 Telescope & \multicolumn{2}{c}{William Herschel 4.2-m} \\
 Detector & \multicolumn{2}{c}{EEV12 2148$\times$4200} \\
 Pixel size & \multicolumn{2}{c}{13.5~$\mu$m} \\
 Efficiency (optics/total) & \multicolumn{2}{c}{$\approx$35\%/14.7\%} \\
\noalign{\smallskip}
\tableline
\end{tabular}
\end{table}

To construct the sample, we first compiled a complete list of ellipticals,
lenticulars, and spiral bulges for which {\tt SAURON} can measure the stellar
kinematics.  Given the specifications of the instrument, this leads to the
following constraints: $-6^\circ \leq \delta \leq 64^\circ$ (zenith distance),
$cz \leq 3000$~km~s$^{-1}$ (spectral range), $M_B \leq -18$ and $\sigma_c \geq
75$~km~s$^{-1}$ (spectral resolution). We further restricted the objects to
$|b| \geq 15^\circ$ to avoid crowded fields and large Galactic extinctions.
All distances are based on a Virgocentric flow model. For galaxies in the
Virgo cluster, Coma~I cloud, and Leo~I group, which we refer to as `cluster'
galaxies, we adopted common mean distances based on the mean heliocentric
velocity of each group (Mould et al.\ 1993). For galaxies outside these three
associations, which we refer to as `field' galaxies, we used individual
distances.

The complete list of galaxies contains 327 objects which we divided into six
categories, first separating `cluster' and `field' galaxies, and then
splitting each of these into E, S0, and Sa bulges. We then selected the {\em
  representative} sample of objects by populating the six resulting
ellipticity versus absolute magnitude planes nearly uniformly. The result is
36 cluster galaxies (12~E, 12~S0, and 12~Sa) and 36 field galaxies (12~E,
12~S0, and 12~Sa), as illustrated in Figure~1. By construction, our sample
covers the full range of environment, flattening, rotational support, nuclear
cusp slope, isophotal shape, etc. It is also large enough to be sub-divided by
any of these criteria, and allow a useful comparison of the sub-samples, yet
small enough that full mapping with {\tt SAURON} is possible over a few
observing seasons.  The 72 galaxies correspond to 22\% of the complete sample
and, as can be seen from Figure~1, remain representative of it. A more
complete description of the sample as well as a listing are available in de
Zeeuw et al.\ (2001, hereafter Paper~II).  Over two-thirds of the sample have
been observed as of September 2001. Completion is expected in April 2002.

\begin{figure}
\plotone{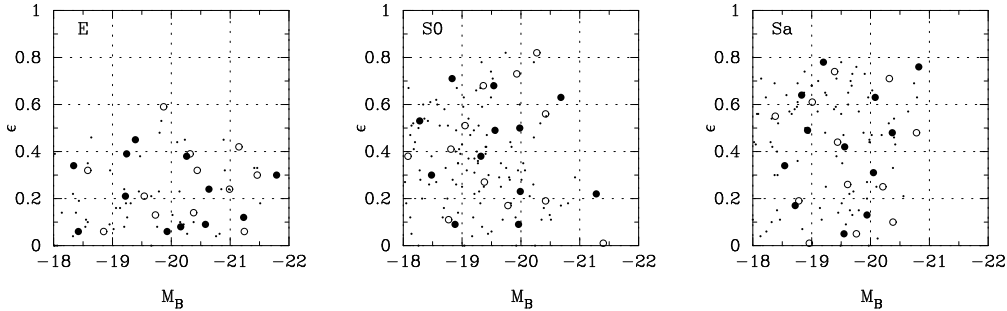}
\caption{Distribution of E, S0, and Sa galaxies in the {\tt SAURON}
  representative sample, in the planes of ellipticity $\varepsilon$ versus
  absolute blue magnitude $M_B$. Open circles: field galaxies. Filled
  circles: cluster galaxies. Small dots: remaining galaxies of the complete
  sample.}
\end{figure}

\section{Stellar Kinematics}

Our strategy is to map galaxies to one effective radius $R_e$, which for
nearly half the sample requires only one pointing. For the largest galaxies,
mosaics of two or three pointings reach 0.5~$R_e$.  Each pointing is split
into four 1800~s exposures dithered by one lenslet.  We reduce the raw {\tt
  SAURON} data as described in Paper~I and derive maps of the stellar
kinematics using the FCQ method (Bender 1990).  This provides the mean stellar
velocity $V$, the velocity dispersion $\sigma$, and the Gauss-Hermite moments
$h_3$ and $h_4$ (e.g.\ van der Marel \& Franx 1993).\looseness=-2

In this section, we present {\tt SAURON} stellar kinematics for three objects
observed in the LR mode. All show the presence of a central stellar disk, with
varying strengths. Other morphologies are illustrated in Papers~I and II.

\subsection{NGC~3384}

NGC~3384 is a large SB0$^-$(s) galaxy in the Leo~I group ($M_B$= --19.6). It
forms a triple on the sky with NGC~3379 and NGC~3389 but there is only
marginal evidence for interactions. The light distribution in the central
$\approx20\arcsec$ is complex. The inner isophotes are elongated along the
major axis, suggesting an embedded disk, but beyond $10\arcsec$ the elongation
is along the minor-axis (e.g.\ Busarello et al.\ 1996). The isophotes are
off-centered at much larger radii. NGC~3384 shows no emission lines, remains
undetected in HI, CO, radio continuum, and X-ray, but has IRAS detections at
12 and 100~$\mu$m (e.g.\ Roberts et al.\ 1991).

Figure~2 displays the stellar kinematics of NGC~3384 and illustrates a key
advantage of {\tt SAURON}. Integrating the flux in wavelength, the surface
brightness distribution of the galaxy is recovered and there is no doubt about
the relative location of the measurements. Figure~2 shows that the bulge of
NGC~3384 is rotating regularly. The mean velocities increase steeply along the
major axis up to $r\approx4\arcsec$, then decrease slightly, and rise again.
No velocity gradient is observed along the minor axis. The velocity dispersion
map shows a symmetric dumb-bell structure and the $h_3$ map is anti-correlated
with $V$ in the inner parts, revealing an abrupt change in the gradient at
$r\approx4\arcsec$ (see also Fisher 1997). All these facts point to the
presence of central (and cold) stellar disk in NGC~3384.

\begin{figure}
\plotone{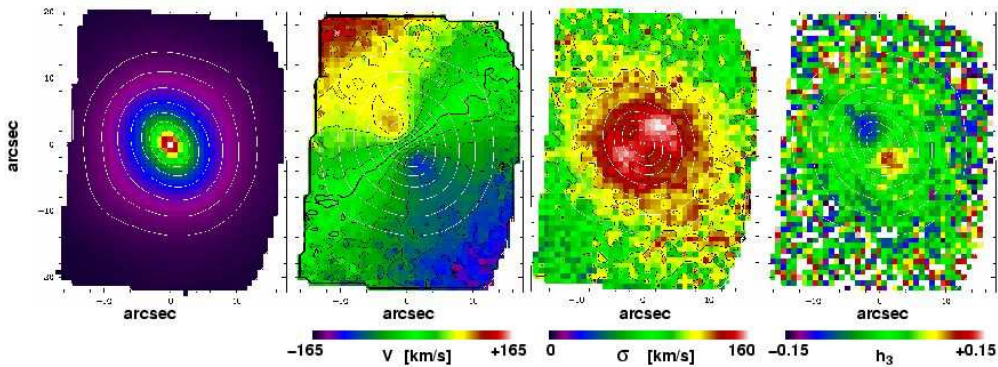}
\caption{{\tt SAURON} absorption-line measurements of the SB0 galaxy
  NGC~3384, based on a single pointing of $4\times 1800$~s. The effective
  spatial sampling is $0\farcs8\times0\farcs8$ and the seeing was
  $\approx2\farcs5$. a) Reconstructed total intensity $I$. b) Mean stellar
  velocity $V$. c) Stellar velocity dispersion $\sigma$. d) Gauss-Hermite
  moment $h_3$. The Gauss-Hermite moment $h_4$ displays little variation over
  the field and is not shown.}
\end{figure}

\subsection{NGC~4526 and NGC~4459}

Many other galaxies in our sample show evidence of a central stellar disk.
NGC~3623 was discussed in Paper~II. Figure~3 shows two other cases where the
stellar disk appears to corotate with a central gaseous disk limited by the
dust lane (Rubin et al.\ 1997).  NGC~4526 is a highly inclined SAB0$^0$(s)
galaxy in the Virgo cluster ($M_B$= --20.7).  The stellar disk is not readily
visible in the reconstructed image, but it is evident in the velocity and
velocity dispersion fields. As in NGC~3384, the rotation along the major axis
first increases, then decreases, and increases again in the outer parts.
However, the extent of the disk is much larger than in NGC~3384. The disk
appears almost edge-on, giving rise to an elongated depression across the
(hot) spheroid in the velocity dispersion map, and completely overwhelming the
central velocity dispersion peak.

NGC~4459 is an S0$^+$(r) galaxy ($M_B$= --20.0) also located in Virgo and
harbouring a $7.3\times10^7$~M$_\odot$ black hole (Sarzi et al.\ 2001). The
same velocity behavior as in NGC~3384 and NGC~4526 is observed along the major
axis (see also Peterson 1978), although the minimum is shallower. The
isovelocity contours are also less skewed, either indicating that the disk is
seen more face-on or that it is intrinsically thicker (or both). This is
supported by the absence of a clear disk signature in the velocity dispersion
map.

\section{Gaseous Kinematics and Ionization Mechanisms}

We now illustrate the scientific potential of the {\tt SAURON} gaseous data.
Paper~II describes how the H$\beta$, [NI], and [OIII] emission lines are
disentangled from the absorption lines by means of a spectral library, and it
summarizes the procedures for deriving fluxes and kinematics. Results on the
non-axisymmetric gaseous disks in NGC~3377 and NGC~5813 are presented in
Papers~I and II, respectively.

\subsection{NGC~7742}

NGC~7742 is a face-on Sb(r) spiral ($M_B$= --19.8) in a binary system. It is
among the latest spirals included in our sample. De Vaucouleurs \& Buta (1980)
identified the inner stellar ring; Pogge \& Eskridge (1993) later detected a
corresponding small, bright ring of HII regions with faint floculent spiral
arms. NGC~7742 possesses a large amount of HI, molecular gas, and dust (e.g.\
Roberts et al.\ 1991) and is classified as a LINER/HII object (Ho, Filippenko,
\& Sargent 1997).

\begin{figure}
\plotone{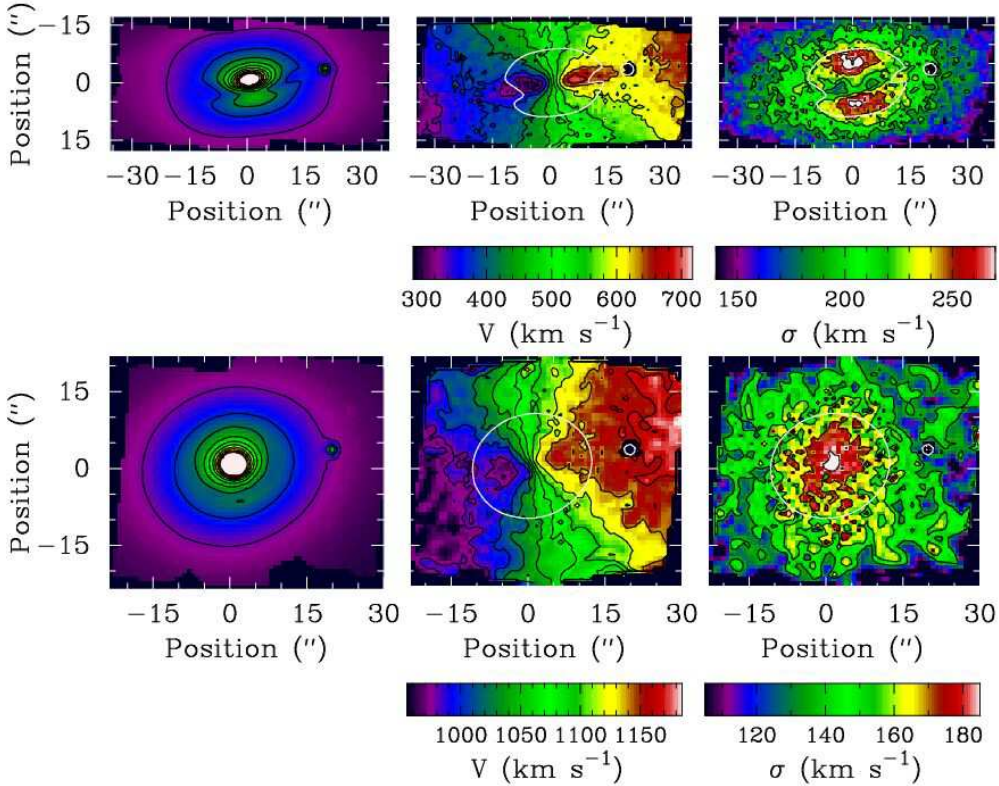}
\caption{{\tt SAURON} absorption-line measurements of the SB0 galaxies
  NGC~4526 and NGC~4459, both based on two partially overlapping pointings.
  Top row: Reconstructed total intensity, mean stellar velocity, and velocity
  dispersion fields of NGC~4526. Bottom row: Same for NGC~4459. Isophotes are
  overlaid on all fields as a reference.}
\end{figure}

Figure~4 shows the [OIII] and H$\beta$ intensity maps, together with the
derived velocity and velocity dispersion fields. Most of the emission is
confined to a ring coinciding with the spiral arms.  H$\beta$ dominates in the
ring (H$\beta$/[OIII]$\approx7-16$) but it is much weaker in the center
(H$\beta$/[OIII]$\approx1$). Also shown in Figure~4 is a reconstructed image
composed of [OIII] and stellar continuum, and a similar image composed of
HST/WFPC2 exposures. The {\tt SAURON} data does not have HST's spatial
resolution, but it does show that our algorithms yield accurate emission-line
maps. The main surprise comes from the stellar and gas kinematics: the gas and
stars within the ring are counter-rotating.

\subsection{NGC~4278}

NGC~4278 is an E1-2 galaxy ($M_B$= --19.9) located in the Virgo cluster. It
contains large-scale dust, as well as a blue central point source (Carollo et
al.\ 1997). Long-slit spectroscopy reveals a peculiar stellar rotation curve,
rising rapidly at small distances from the nucleus and dropping to nearly zero
beyond $\approx30\arcsec$ (Davies \& Birkinshaw 1988; van der Marel \& Franx
1993). NGC 4278 also contains a massive HI disk extending beyond 10~$R_e$. The
HI velocity field is regular but has non-perpendicular kinematic axes,
indicating non-circular motions (Raimond et al\ 1981; Lees 1992).\looseness=-2

Figure~5 displays the reconstructed stellar intensity as well as the [OIII]
map derived from two {\tt SAURON} pointings. Despite the regular and
well-aligned stellar isophotes, the distribution of ionized gas is very
extended and strongly non-axisymmetric. Its shape is reminiscent of a bar
terminated by ansae, as observed in spiral galaxies. It will be interesting to
construct a comprehensive dynamical model for NGC~4278, and explore its
orbital structure and dark matter content.

\begin{figure}
\plotone{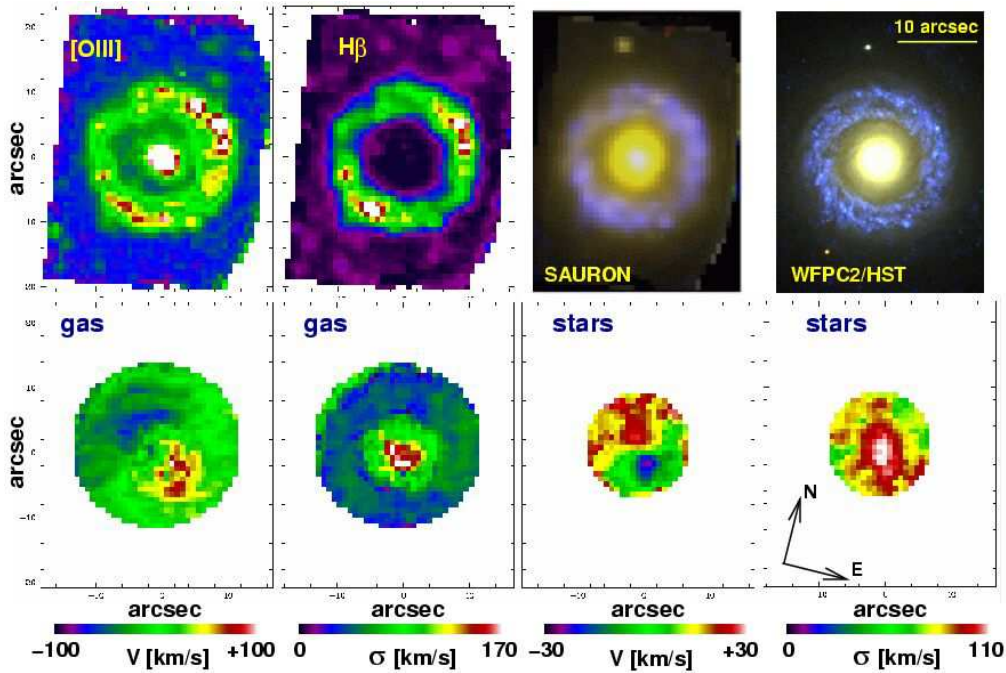}
\caption{{\tt SAURON} measurements of the stars and gas in NGC~7742,
  based on one pointing with seeing 1.5-2\farcs5. The top row shows the
  emission-line intensity distributions of [OIII] and H$\beta$, followed by a
  reconstructed image composed of [OIII] and stellar continuum and a similar
  HST/WFPC2 image composed of F336W, F555W, and F814W exposures. The bottom
  row shows the derived gas velocity and velocity dispersion fields, followed
  by the stellar velocity and velocity dispersion.}
\end{figure}

\begin{figure}
{\centering\leavevmode\epsfxsize=0.9\textwidth\epsfbox{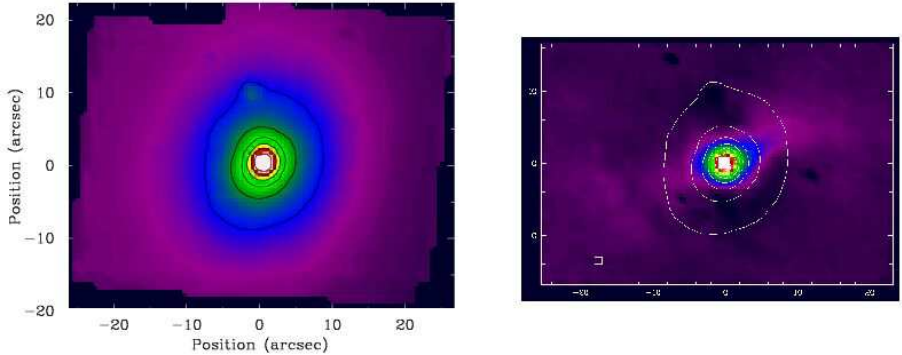}}
\caption{{\tt SAURON} measurements of the stars and gas in NGC~4278,
  based on two partially overlapping pointings. The reconstructed stellar
  intensity is shown on the left and the [OIII] intensity on the right.
  Isophotes are overlaid on both images. The scales are identical but the
  fields-of-view differ slightly. The projected size of a {\tt SAURON} lenslet
  is indicated in the bottom left corner of the second image.}
\end{figure}

\section{Stellar Populations}

{\tt SAURON}'s wavelength range allows two-dimensional mapping of the
line-strength indices H$\beta$, Mg\,{\it b}, and Fe5270. To convert measured
equivalent widths to indices on the Lick/IDS system (Worthey et al.\ 1994),
corrections for the difference in spectral resolution and velocity broadening
in the galaxies must be applied. Both are described in Paper~II. With stellar
population models, these indices can be used to estimate luminosity-weighted
ages and metallicities and study their contours. We discuss below the stellar
populations of galaxies with kinematically decoupled cores. NGC~3384 shows a
similar behaviour (Paper~II).

Davies et al.\ (2001) discussed the {\tt SAURON} kinematics and line-strengths
of the large E3 galaxy NGC~4365 ($M_B$= --20.9) in the Virgo cluster. While
the center and main body of the galaxy are decoupled kinematically, both
components show the same luminosity-weighted age ($\approx14$~Gyr) and a
smooth metallicity gradient is observed, suggesting formation through gas-rich
mergers at high redshift.\looseness=-2

\subsection{NGC~5813}

NGC~5813 is an E1--2 galaxy in the Virgo-Libra Cloud ($M_B= -21.0$).  It is
undetected in HI or CO but has an unresolved, weak central radio continuum
source and emission-line ratios typical of LINERS (Birkinshaw \& Davies 1985;
Ho, Filippenko, \& Sargent 1997).  The ionized gas exhibits a complex
filamentary structure most likely not (yet) in equilibrium (Caon et al.\
2000; Paper~II).

As illustrated in Figure~6, NGC~5813 harbours a kinematically decoupled core
(see also Efstathiou, Ellis, \& Carter 1982). While the body of the galaxy
appears purely pressure supported, the center rotates rapidly around an axis
tilted by $\approx13^\circ$ from the photometric minor axis. As in NGC~4365,
the H$\beta$ map is featureless, indicating a roughly constant
luminosity-weighted age, but the Mg\,{\it b} and Fe5270 maps show a steep
central gradient, indicating a strong metallicity gradient (see also Gorgas,
Efstathiou, \& Aragon-Salamanca 1990).

\begin{figure}
\plotone{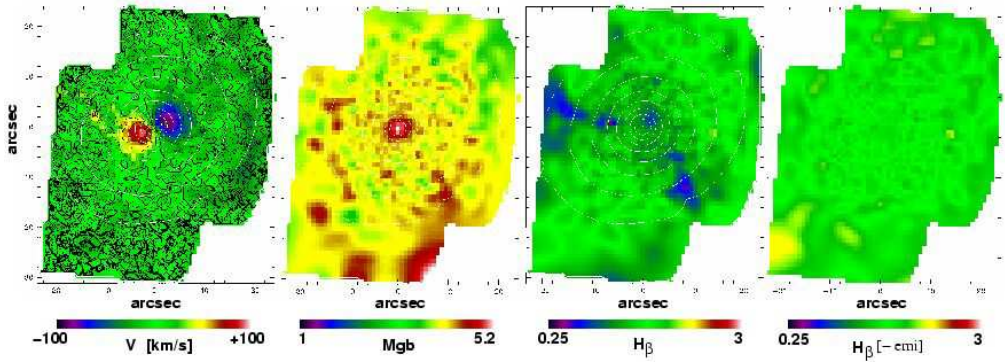}
\caption{{\tt SAURON} maps of NGC~5813, based on two partially
  overlapping pointings with seeing $1-2\farcs5$. From left to right: Stellar
  (mean) velocity field, Mg~{\it b} map, raw H$\beta$ map, and H$\beta$ map
  corrected for emission. Isophotes are superposed on the first three images.}
\end{figure}

\section{Concluding remarks}

The {\tt SAURON} survey will be completed in 2002. The first results show that
early-type galaxies display line-strength distributions and kinematic
structures which are more varied than often assumed.  The sample contains
specific examples of minor axis rotation, decoupled cores, central stellar
disks, and non-axisymmetric and counter-rotating gaseous disks. The
provisional indication is that only a small fraction of these galaxies can
have axisymmetric intrinsic shapes.\looseness=-2

We are complementing the {\tt SAURON} maps with high-spatial-resolution
spectroscopy of the nuclear regions using {\tt OASIS} on the CFHT. {\tt STIS}
spectroscopy for many of the galaxies is in the HST archive. Radial velocities
of planetary nebulae and/or globular clusters in the outer regions have been
obtained for some of the galaxies, and many more will become available to
$\approx$5$R_e$ with a special-purpose instrument now under construction
(Freeman et al.\ 2001, in prep).\looseness=-2

\smallskip
Understanding the formation, structure, and evolution of galaxies is one of
the central drivers in Ken Freeman's research. His enthusiasm and guidance at
all levels is an inspiration to this entire field, and in particular to our
team. We wish him and Margaret all the best for a happy and exciting future.

\smallskip

\acknowledgments It is a pleasure to thank the ING staff for enthusiastic and
competent support on La Palma.  The {\tt SAURON} project is made possible
through grants 614.13.003 and 781.74.203 from ASTRON/NWO and financial
contributions from the Institut National des Sciences de l'Univers, the
Universit\'e Claude Bernard Lyon I, the universities of Durham and Leiden, the
British Council, PPARC grant `Extragalactic Astronomy \& Cosmology at Durham
1998--2002', and the Netherlands Research School for Astronomy NOVA.

\end{document}